\documentclass[11pt]{article}

\usepackage[preprint]{acl}

\usepackage{times}
\usepackage{latexsym}

\usepackage[T1]{fontenc}

\usepackage[utf8]{inputenc}

\usepackage{microtype}

\usepackage{inconsolata}

\usepackage{graphicx}
\usepackage{amsmath}

\usepackage{amssymb}

\newtheorem{theorem}{Theorem}[section]

\newtheorem{definition}[theorem]{Definition}

\usepackage{multirow}
\usepackage{graphicx}
\usepackage{makecell}

\title{Argus: Reorchestrating Static Analysis via a Multi-Agent Ensemble for Full-Chain Security Vulnerability Detection}

\author{
 \textbf{Zi Liang\textsuperscript{1}}\thanks{Equal contributions.},
  \textbf{Qipeng Xie\textsuperscript{2}}\footnotemark[1],
  \textbf{Jun He\textsuperscript{3}}\thanks{Corresponding authors.},
  \textbf{Bohuan Xue\textsuperscript{3}},
  \\
  \textbf{Weizheng Wang\textsuperscript{1}},
  \textbf{Yuandao Cai\textsuperscript{2}},
  \textbf{Fei Luo\textsuperscript{4}},
  \textbf{Boxian Zhang\textsuperscript{3}},
  \textbf{Haibo Hu\textsuperscript{1}}\footnotemark[2],
  \textbf{Kaishun Wu\textsuperscript{2}}
\\
\\
  \textsuperscript{1}The Hong Kong Polytechnic University,
  \textsuperscript{2}HKUST
  \\
  \textsuperscript{3}SF Express,
  \textsuperscript{4} Great Bay University
\\
  \small{
    \textbf{Correspondence:} \href{mailto:hejun@sf-express.com}{hejun@sf-express.com} \href{mailto:haibo.hu@polyu.edu.hk}{haibo.hu@polyu.edu.hk}
  }
}

\begin{document}
\maketitle
\begin{abstract}

Recent advancements in Large Language Models (LLMs) have sparked interest in their application to Static Application Security Testing (SAST), primarily due to their superior contextual reasoning capabilities compared to traditional symbolic or rule-based methods. 
However, existing LLM-based approaches typically attempt to replace human experts directly without integrating effectively with existing SAST tools. 
This lack of integration results in ineffectiveness, including high rates of false positives, hallucinations, limited reasoning depth, and excessive token usage, making them impractical for industrial deployment. 
To overcome these limitations, we present a paradigm shift that reorchestrates the SAST workflow from current LLM-assisted structure to a new LLM-centered workflow. 
We introduce Argus (Agentic and Retrieval-Augmented Guarding System), the first multi-agent framework designed specifically for vulnerability detection. 
Argus incorporates three key novelties: comprehensive supply chain analysis, collaborative multi-agent workflows, and the integration of state-of-the-art techniques such as Retrieval-Augmented Generation (RAG) and ReAct to minimize hallucinations and enhance reasoning. 
Extensive empirical evaluation demonstrates that Argus significantly outperforms existing methods by detecting a higher volume of true vulnerabilities while simultaneously reducing false positives and operational costs. Notably, Argus has identified several critical zero-day vulnerabilities with CVE assignments.
\end{abstract}

\section{Introduction}\label{sec:intro}

The integrity of modern digital ecosystems is under critical threat
from the escalating proliferation of software vulnerabilities. In
2025, more than 50,000 Common Vulnerabilities and Exposures (CVEs)
were disclosed, marking a sharp rise from the previous year and
signaling a rapidly expanding attack surface~\cite{Cves}. These
vulnerabilities result in catastrophic consequences for the financial
sector and society, as exemplified by global outbreaks such as
NotPetya~\cite{aidan2017comprehensive}, the Equifax
breach~\cite{zou2018ve}, React2Shell~\cite{r2s}, and Cloudflare,
which inflicted almost 180 billion USD in damages or compromised the
personal data of over 143 million consumers.

Static Application Security Testing (SAST), as one of the most
effective strategies for vulnerability detection, has thus emerged as
a cornerstone of secure software
development~\cite{li2023comparison}. Many state-of-the-art SAST tools,
including CodeQL~\cite{Codeql}, Infer~\cite{infer}, and Snyk
Code~\cite{snyk}, rely on static taint analysis to trace untrusted
data flows from sources to potentially dangerous sinks. While these
taint analyses can effectively capture specific and well-known
vulnerability patterns with handcrafted symbolic
rules~\cite{avgustinov2016ql}, they lack the flexibility to detect
novel or system-specific flaws~\cite{hin2022linevd,
  kang2022detecting}, and remain fundamentally constrained by
challenges such as high false positive rates, limited scalability in
complex codebases, and insufficient semantic context. More critically,
they often miss real vulnerabilities due to incomplete coverage of
sources and sinks as well as cracked data flows caused by
insufficient support for advanced language features, which together
undermine their practical effectiveness~\cite{BlackHat}.

Consequently, LLM-enhanced vulnerability detection has been
proposed~\cite{xia2022less,lemieux2023codamosa,li2024enhancing,bugbot,rabbit,iris,cve-reproduce,liu2025swell,guo2025repoaudit}. These
methods regard LLMs as vulnerability analysis experts and incorporate
them with SAST tools (e.g., CodeQL) for more accurate sink discovery
and flow construction. While they achieve significant improvements
compared with traditional SAST methods, current incorporations between
LLMs and SAST tools have three significant
challenges. \textbf{First}, these methods remain limited by classical
pipeline-based symbolic analysis in which the reasoning depth is
restricted to single-pass inference, leading to the inability to mining new vulnerabilities. \textbf{Second}, the
introduction of LLMs brings the risk of hallucinations, which further
leads to false positives. 
\textbf{Third}, current straightforward combinations of LLMs
with SAST tools fail to address reachability challenges
such as disrupted data flows. 
As a result, ample
evidence~\cite{zhou2024large,khare2025understanding,guo2025repoaudit}
suggests that these methods still struggle when faced with novel
vulnerability types and restricted contexts, and typically perform
poorly for practical vulnerability detection tasks, which leaves a
question: \emph{how to design the LLM-SAST-tools incorporation that
  can fully employ the potential of LLMs in vulnerability detection?}

To answer this question, we explore the possibility of proposing a new
SAST framework with LLM agents as the first-class citizens. In
contrast to previous studies, which are \emph{SAST-tools-centered} and
\emph{LLM-assisted}, we present a new framework that is
\emph{LLM-centered}, where all other factors, including SAST tools,
code repository dependencies, and existing issue descriptions, serve
as supplemental information for the LLM agent system.

Following this paradigm, we propose Argus (\underline{A}gentic and
\underline{R}etrieval-augmented \underline{G}uarding
\underline{S}ystem), a novel SAST framework for automatic vulnerability detection. Argus reorchestrates the SAST
workflow with the following three significant improvements: \emph{i) Full
  Supply Chain Analysis.} Instead of regarding the code repository as
an ``isolated silo'', we combine it with its dependencies for a more
practical vulnerability evaluation. \emph{ii) Multi-Agent
  Collaboration.} We decouple the SAST pipeline into different modules
and design various agents for them, including dependency scanning,
information collecting, PoC (Proof of Concept) generating, data flow
scanning, and data flow reviewing. All these agents provide a more
elegant implementation of vulnerability detection. \emph{iii)
  Introducing New Techniques.} We introduce the state-of-the-art agentic
strategies to address the difficulties of LLM-based SAST analysis. For
instance, we apply RAG (Retrieval-Augmented Generation) to
collect community vulnerability information from dependencies and code
repositories, enabling more comprehensive and accurate reports and
sink identification. We utilize ReAct~\cite{yao2023react} to meet the long-term
reasoning requirements in data flow review and PoC generation. Through
this first-time integrated design, Argus preserves the precision of
symbolic analysis while harnessing the deep contextual reasoning of
LLMs, enabling a more general, adaptable, and robust approach to
identifying sophisticated vulnerabilities in real-world software.

Our contributions can be summarized as follows:
\begin{itemize}
\item We propose Argus, a new framework for static code vulnerability
  analysis. To the best of our knowledge, it is the first LLM-centered
  SAST framework. It is also the first framework that considers the
  supply chain and introduces ReAct and RAG strategies.
\item We comprehensively evaluate Argus on popular industry-level
  open-source code repositories with tens of millions of lines of
  source code and tens of thousands of GitHub stars. Extensive
  empirical results show that Argus exhibits lower token consumption
  than existing methods and can mine more new vulnerabilities.
  Using Argus, we discovered potential zero-day vulnerabilities in these thoroughly evaluated codebases and these vulnerabilities have been assigned with CVE numbers.
\item We will release our framework as a tool to benefit the security
  of industry and academia.
\end{itemize}


\section{Related Work}

\textbf{LLM-based Vulnerability Detection \& Discovery.}
Recent
studies~\cite{cve-reproduce,liu2025swell,iris,guo2025repoaudit}
have demonstrated that LLM agents are increasingly promising for
achieving more precise automatic vulnerability discovery. On the one
hand, preliminary research shows that LLMs can directly identify
common vulnerability patterns (e.g., CWE categories) from source code,
especially when augmented with contextual metadata such as commit
messages or issue descriptions~\cite{guo2025repoaudit}. On the other
hand, LLMs can be employed to mine unknown sources or sinks within the
source code, enabling more accurate static taint analysis for
vulnerability detection~\cite{iris,guo2025repoaudit}. Systematic
evaluations have also been provided~\cite{llbezpeky} suggesting that
even prompt-engineering-based LLMs can surprisingly improve the
detection rate for Android vulnerabilities. However, as highlighted by
recent literature reviews~\cite{zhou2024large,khare2025understanding,zhou2025large}, while LLMs show
promise in both detection and repair, their performance remains highly
sensitive to input formulation, code context length, and the quality
of labeled vulnerability data.

\begin{figure}
\includegraphics[width=\linewidth]{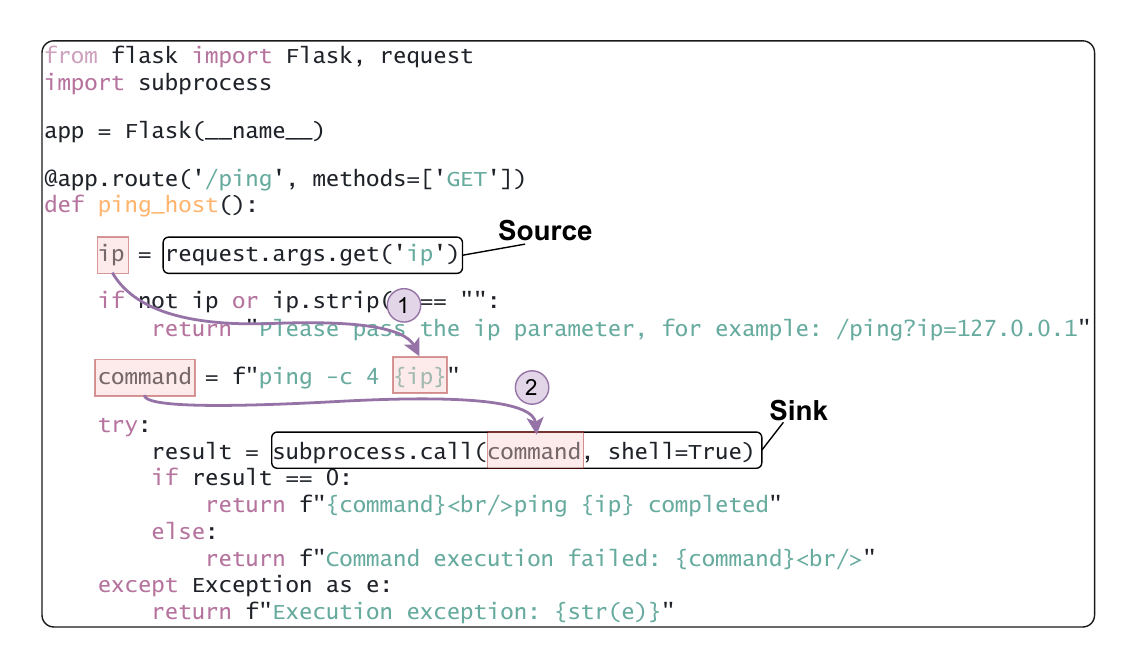}
\caption{A toy Python example of vulnerability detection. The function \texttt{subprocess.call($\cdot$, shell=True)} acts as the \emph{sink}, which may receive injected malicious input from the \emph{source} \texttt{ip} and cause it to be executed. Data flow of this vulnerability is represented by the connecting purple arrows.}
\label{fig:vul}
\end{figure}

\textbf{Techniques of LLM Agents.}
Recent research on LLM-based agents has emphasized structured
reasoning~\cite{yao2023react,rea2,rea3}, tool integration, and memory
mechanisms to support complex task execution. The ReAct
framework~\cite{yao2023react}, which interleaves reasoning steps with
action executions (e.g., API calls), has become a standard
architecture for agentic behavior, enabling agents to dynamically
reason and reflect on observations. This paradigm has
been extended through retrieval-augmented generation
(RAG)~\cite{lewis2024rag,rag2,rag3}, allowing agents to access
external knowledge bases during inference to ground responses in
up-to-date or domain-specific information. More recent
works~\cite{wang2025planner} explore modular agent designs. For
instance, \citet{wang2025planner} proposes a planner-executor
architecture that separates high-level task decomposition from
low-level tool invocation, significantly improving success rates on
multi-step reasoning benchmarks. Additionally, empirical studies have
shown that chain-of-thought (CoT)~\cite{cot} prompting and its
variants (e.g., self-reflection, step-back prompting) consistently
enhance agent reliability across diverse
environments~\cite{zhu2025llm}. These developments highlight a
methodological shift from monolithic prompting toward compositional,
tool-augmented agent systems that balance reasoning fidelity with
external interaction.



\section{Methodology}

\begin{figure*}
\includegraphics[width=\linewidth]{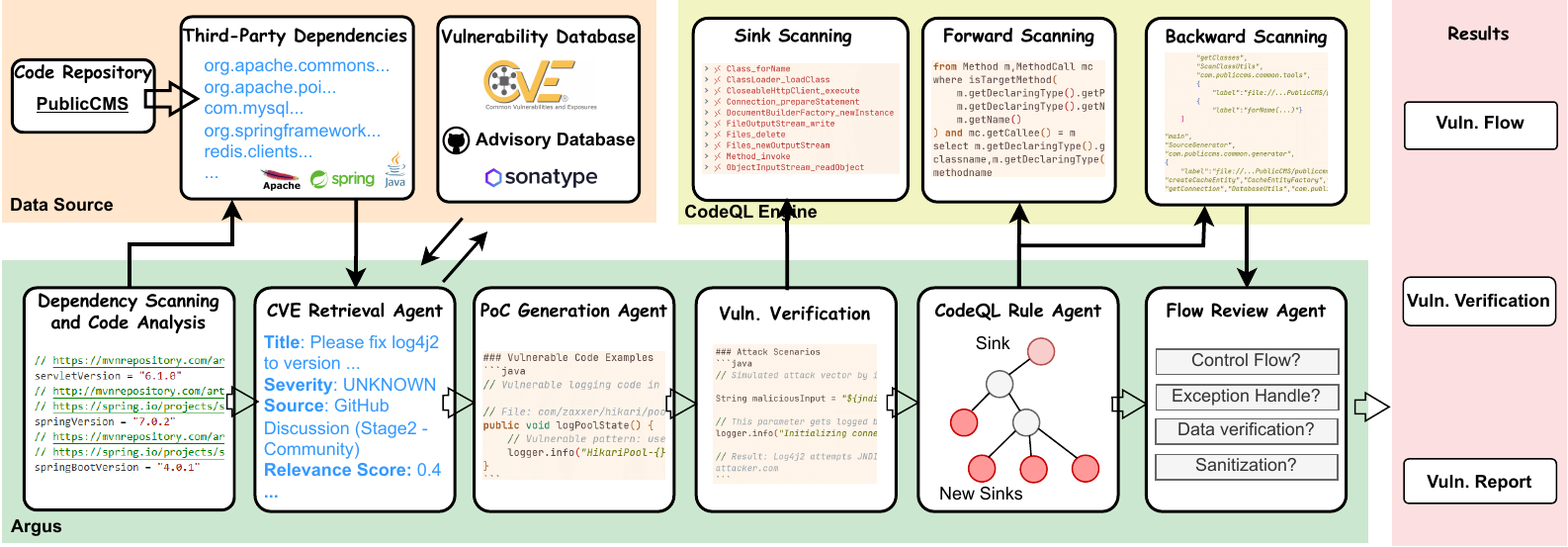}
\caption{The overall workflow of the Argus framework. Given a code
  repository, Argus first parses it with its third-party dependencies and
  retrieves potential issues or vulnerabilities associated with
  them. For suspected vulnerabilities, Argus generates proofs of
  concept (PoCs) as well as corresponding repairs. It then integrates
  with CodeQL for sink identification and employs our \emph{Re}$^3$
  (Retrieval, Recursion, and Review) method, which combines forward
  and backward scanning to achieve comprehensive data flow
  extraction. Finally, the framework produces a detailed report
  containing all detected vulnerabilities, along with their data flows
  and verification details, for human review. All information shown in
  this figure is derived from actual execution results of Argus.}
\label{fig:argus}
\end{figure*}

\subsection{Problem Definition}
In this section, we begin with the fundamentals of current static
analysis for security vulnerability detection. A toy example of these
concepts is shown in Figure \ref{fig:vul}.

\noindent
\textbf{Taint \& Content.}
A \emph{taint} refers to a variable or object that stores unsafe or injected
content occurring in the code. The taint is typically held within a
variable or container of the program, which we refer to as a \emph{content
node}. A content node can be a variable, an array, a list, a key or
value in a hashmap, or a member of an object. Considering the
properties of these content nodes, we emphasize two special nodes for
vulnerability detection:
\emph{i) \textbf{Sources}}, which indicate unauthorized user interface
inputs in the source code that might be maliciously exploited by an
adversary for security attacks. These can include HTTP request
parameters, text or files uploaded by users, or inputs from
environment variables and databases.
\emph{ii) \textbf{Sinks}}, which refer to executed functions in the
source code that might involve dangerous or sensitive operations, such
as SQL executions or system command executions.

\noindent
\textbf{Access Paths.}
An access path records how content propagates between different node types.

\begin{definition}[Data Flow]\label{def:df}
  Given a source code project $\mathcal{S}=\{S_{i}\}_{i=1,\dots,N_{S}}$ consisting of $N_{S}$ source code files, where each file $S_{i}$ is a text string, let $\mathbb{V}_{S}$ and $\mathbb{E}_{S}$ denote the sets of content nodes and access paths in $\mathcal{S}$, respectively. A data flow from a source node to a sink node is defined as a sequence of logical edge triples:
  \begin{equation}
    \label{eq:df}
    F_{S}^{(v_{i}^{1},v_{k}^{n})} = \left[ (v_{i}^{1}, e_{j}^{1}, v_{k}^{1})_1, \dots, (v_{i}^{n}, e_{j}^{n}, v_{k}^{n})_n \right],
  \end{equation}
  where $n < N_{F}$ is the length of the flow, $v_{i}^{m}, v_{k}^{m} \in \mathbb{V}_{S}$ and $e_{j}^{m} \in \mathbb{E}_{S}$ for all $m \in \{1,\dots,n\}$, and the flow satisfies the continuity condition $v_{k}^{m-1} = v_{i}^{m}$ for $m=2,\dots,n$. The data flow graph of $\mathcal{S}$ is the set of all such data flows, i.e., $\mathcal{F} = \{ F_{S}^{(v_{i}^{1},v_{k}^{n})} \mid v_{i}^{1}, v_{k}^{n} \in \mathbb{V}_{S} \}$.
\end{definition}

Given Definition~\ref{def:df} and the notions of sources and sinks, we
can define the vulnerability and the corresponding detection task as
follows:

\begin{definition}[Vulnerability Detection]\label{def:vd}
Given a source node $v_{so} \in \mathcal{V}_{so}$ and a sink node $v_{si} \in \mathcal{V}_{si}$, a vulnerability in $\mathcal{S}$ is defined as a reachable data flow from $v_{so}$ to $v_{si}$, denoted as $F_{S}^{(v_{so},v_{si})}$.  
Correspondingly, the vulnerability detection task consists of two steps: \textbf{i)} constructing the source set $\mathcal{V}_{so}$ and the sink set $\mathcal{V}_{si}$, and \textbf{ii)} mining all reachable data flows between them to obtain the vulnerability set:
\begin{equation}
  \label{eq:vul}
  \mathcal{F}_{S}^{vul} = \{ F_{S}^{(v_{so},v_{si})} \mid v_{so} \in \mathcal{V}_{so},\ v_{si} \in \mathcal{V}_{si} \}.
\end{equation}
\end{definition}

Under the research scope of Definition~\ref{def:vd}, current SAST
methods employ cybersecurity databases to construct source and sink
sets and rely on rule-based engines (e.g., CodeQL) for automatic data
flow discovery. Although LLMs have been employed in some steps of this
task (e.g., sink set discovery), a unified, agentic, and in-depth
integration of LLMs into vulnerability detection remains largely
uninvestigated, which significantly limits the potential of
AI-enhanced vulnerability discovery.
Can LLMs effectively reveal existing CVEs? Are they able to mine
zero-day vulnerabilities? How can we mitigate the inherent drawbacks
of both LLMs and traditional SAST tools? We aim to answer these
questions by proposing a new framework in the following subsections.


\begin{figure}[t]
\includegraphics[width=\linewidth]{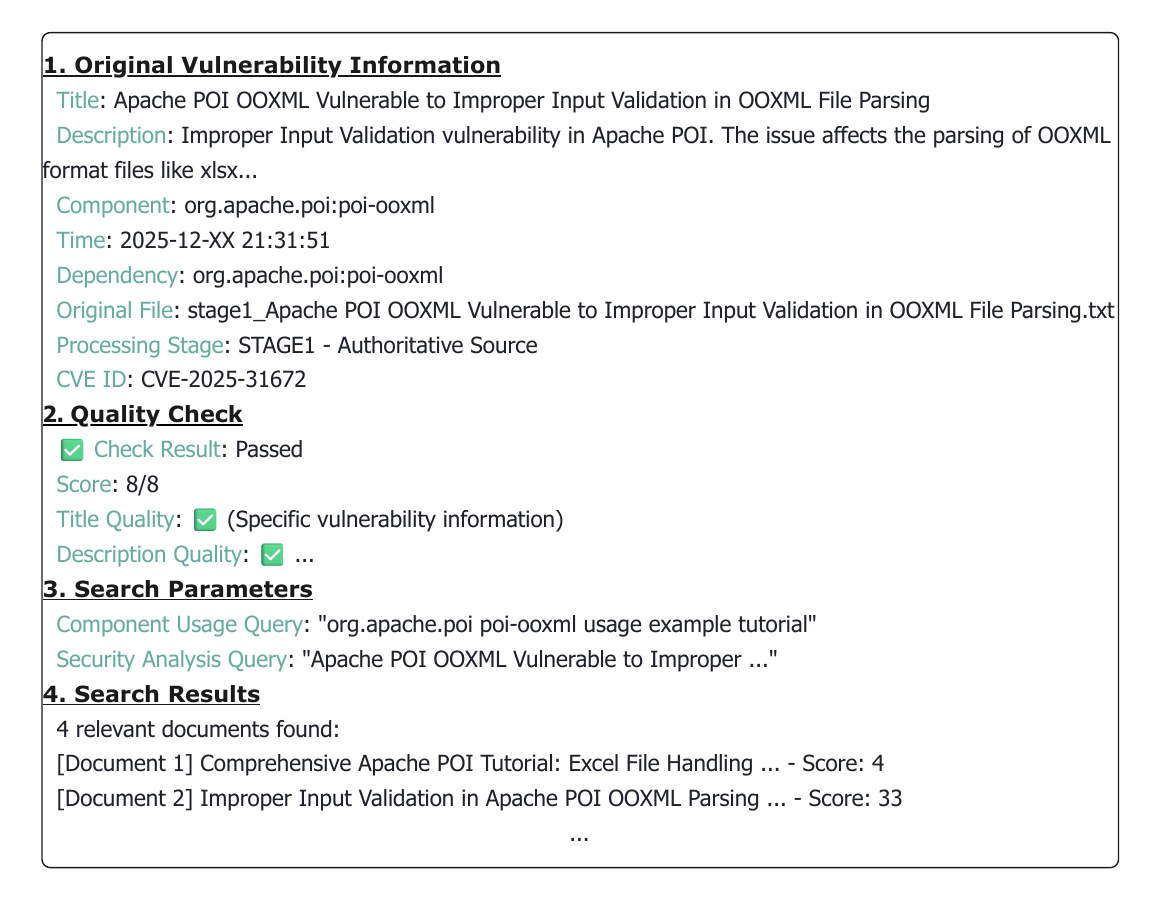}
\caption{A practical example of the structured information summerized
  by our RAG agent. It summarizes all necessary information for the
  locating of potential sinks given a target dependency (e.g., \texttt{org.apache.poi:poi-ooxml}).}
\label{fig:vulns-search}
\end{figure}

\begin{figure}[t]
\includegraphics[width=\linewidth]{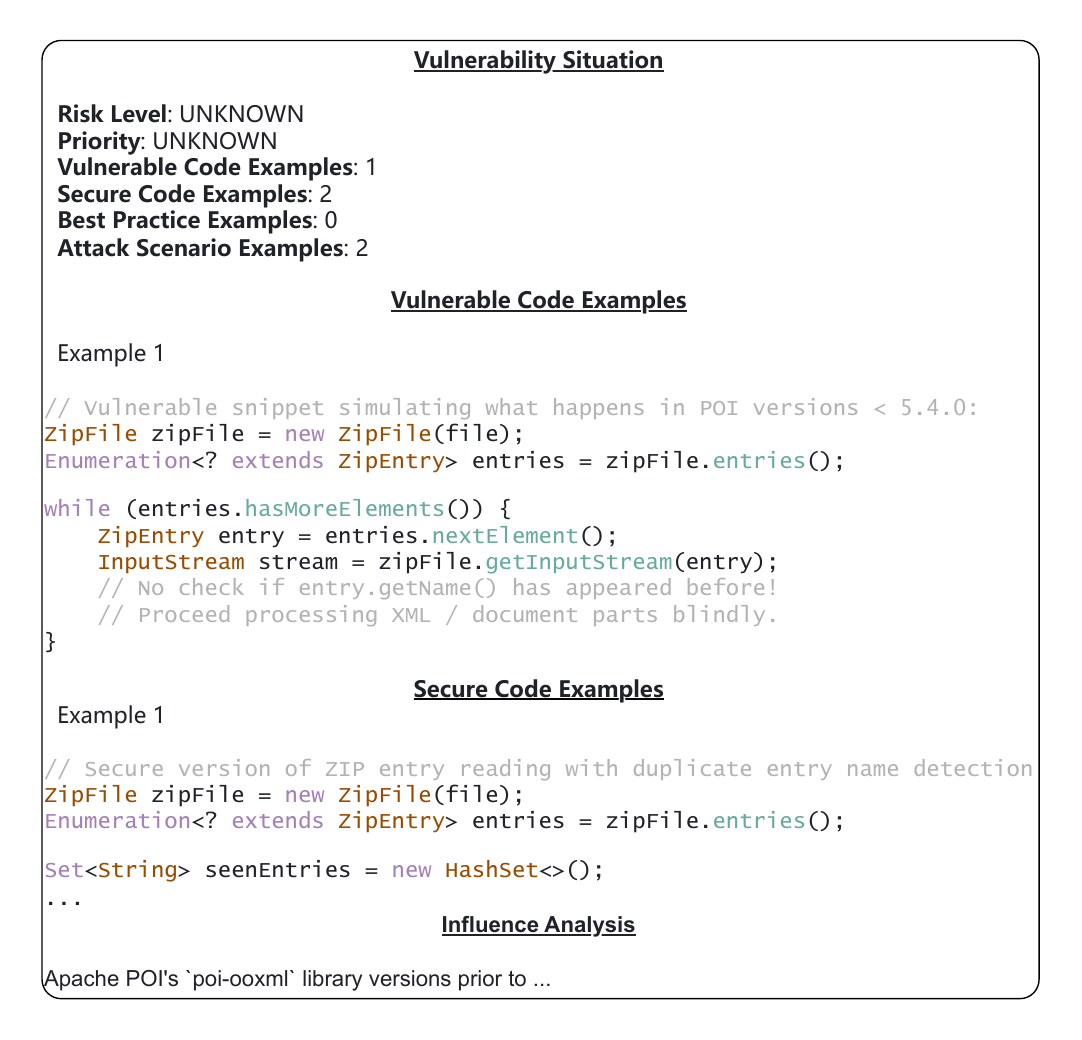}
\caption{Example of PoC generation and vulnerability verification. The
POC agent will think about the possibility of writing proof-of-concept
example code which can successfully exploit the codebase. In this way,
we can ensure the correctness of sinks and start our data flow
analysis in the next step.}
\label{fig:poc-example}
\end{figure}

\subsection{Argus: Agentic and Retrieval-Augmented Guarding System}
As shown in Figure~\ref{fig:argus}, the Argus framework consists of
two core components: \emph{RAG-enhanced full supply chain sink
  analysis} and \emph{Re$^3$-based data flow analysis}. Both
components tightly integrate LLMs with existing vulnerability
resources (e.g., NVD) and static analysis tools (e.g., CodeQL),
resulting in more effective and efficient vulnerability detection.

\subsubsection{Sinks Scanning: RAG-Enhanced Full Supply Chain Analysis}

The first step in identifying vulnerable data flows is sink
scanning. Once sinks are identified, we can search for all potential
unsafe data flow candidates. Unfortunately, current vulnerability
detection
strategies~\cite{cve-reproduce,liu2025swell,iris}
exhibit two significant drawbacks: \emph{i)} they directly employ LLMs
for sink discovery, resulting in low scanning efficiency, low
precision, and low recall, which significantly impacts the
effectiveness of subsequent vulnerability detection steps; \emph{ii)}
they focus merely on the security of the project source code
$\mathcal{S}$ itself, severely neglecting its dependencies (i.e., the
supply chain). However, these dependencies across different versions
may contain known vulnerabilities or expose new, unexpected
vulnerabilities or sinks due to improper sanitization. These drawbacks
render automatic scanning unsuitable for industrial-level
vulnerability detection.

To address these issues and facilitate subsequent static analysis for
vulnerability detection, we design a dedicated component that parses
project dependencies and performs retrieval-augmented generation to
mine all relevant vulnerability types.

\noindent
\textbf{Dependencies Parsing.} Argus first parses the project
management file (e.g., \texttt{pom.xml} in Java projects) to collect
all active dependencies specified in the configuration files. It then
looks up their actual usage in the source code. The detailed
information of these dependencies is passed to the next step for
further analysis.

\noindent
\textbf{Retrieval.} Given the identified dependencies, RAG agents in
Argus perform parallel searches for vulnerability information from the
Internet. We design two types of search resources for Argus: \emph{i)
  Authoritative Sources.} We select the National Vulnerability
Database (NVD), Open Source Vulnerabilities (OSV), GitHub Security
Advisories (GHSA), and Snyk Vulnerability Database (Snyk) as official
vulnerability sources. For each dependency, we query these databases
and extract vulnerability information into structured objects (as
detailed in Figure~\ref{fig:vulns-search}), including descriptions,
severities, affected versions, and CVE IDs (if available). \emph{ii)
  Community Sources.} We employ a hierarchical approach to collect
community-reported vulnerability information for the target
dependency. Specifically, we examine vulnerability-related issues from
both the dependency's primary repository and its derived (forked)
repositories, prioritizing those explicitly linked to CVEs or GHSAs.
We define three indicators to assess the quality of retrieved
vulnerabilities: relevance score, credibility score, and the content
    quality score, as detailed in Appendix \ref{sec:impl-detail}.

\noindent
\textbf{Proof of Concepts (PoC) of Vulnerabilities.}
Given the retrieved vulnerabilities, Argus employs a dedicated PoC
agent to iteratively verify their exploitability. Specifically, the
PoC agent follows the ReAct paradigm~\cite{yao2023react}: it first prompts the
LLM to re-describe the vulnerability details, then sequentially
reasons about the root cause, code patterns, and potential attack
scenarios. Subsequently, the agent generates source code that triggers
the vulnerability, along with a corresponding patch and additional
explanatory information. Figure~\ref{fig:poc-example} provides a toy
example of this process.
In this way, Argus enables effective and efficient scanning of known
vulnerabilities in a code repository, complete with verifiable
PoCs. Once verified, the resulting vulnerabilities, repairs, and detailed reports can
be directly applied to the repository, offering greater
comprehensiveness than previous vulnerability reproduction
methods~\cite{cve-reproduce}.
Moreover, these verified vulnerabilities and their PoCs serve as
valuable context to inspire Argus in discovering additional
vulnerabilities, including zero-day ones, as detailed in the next
section.

\subsubsection{Data Flow Analysis: Retrieval, Recursion, and Review}
Given candidate vulnerability descriptions and their corresponding
sinks, the final step in vulnerability detection is to recover and
construct the complete data flow graph that reflects how taint
propagates from sources to these sinks. A path is confirmed as a
verified vulnerability if it reaches a sink without encountering
effective sanitization.
However, tracking data flows to sinks suffers from the well-known
\emph{stealthy interruptions problem}: mainstream static analysis
tools, such as CodeQL, often fail to resolve data connections arising
from advanced language features, including \emph{reflection}, \emph{multi-thread
interactions}, and \emph{pointer aliasing}. Unfortunately, directly applying
LLMs to this task provides limited benefit, as the complexity of data
flow analysis for each sink far exceeds the reasoning capacity of a
single LLM inference. To address this challenge, we propose a novel
workflow named \emph{Re}$^3$ (\underline{Re}trieval, \underline{Re}cursion,
and \underline{Re}view) that enables
more accurate and robust data flow analysis.

\noindent
$\bullet~$\textbf{Retrieval.}
Similar to existing LLM-based vulnerability detection
approaches~\cite{iris,liu2025swell}, we employ CodeQL to
search for all candidate paths from sources to each targeted sink. If
data flows to the sink are successfully identified, they are stored
for the subsequent \textbf{Review} step. Otherwise, the sink is
forwarded to the \textbf{Recursion} step.

\noindent
$\bullet~$\textbf{Recursion.}
For sinks that cannot be reached through direct forward analysis, we
introduce a backward-forward mechanism to bridge
interruptions. Specifically, we iteratively analyze the upstream
function calls of the sink until the reverse traversal
terminates. This process constructs a tree rooted at the original
sink, with leaves corresponding to the uppermost function calls. We
then treat these leaves as new surrogate sinks and perform forward
data flow analysis using CodeQL. The resulting forward flows are
connected to the backward tree paths to form candidate vulnerable data
flows. Note that not all data flows produced in this step are
guaranteed to be valid, necessitating a final \textbf{Review} step.

\noindent
$\bullet~$\textbf{Review.}
At the end of the data flow analysis, we employ a dedicated LLM agent
to audit the correctness of candidate vulnerabilities through three
sequential steps: \emph{i) End-to-End Reachability Analysis}, in which
the LLM is prompted to determine whether control flow structures
(e.g., \texttt{if/else}, \texttt{switch}), exception handling (e.g.,
\texttt{try/catch}), or data validation mechanisms interrupt or alter
the original data flow to the sink. \emph{ii) Hop-by-Hop Analysis},
which conducts a more detailed examination of the entire data flow,
verifying at each step how the taint enters a code block, the content
and access path involved, and most importantly, whether any form of
input validation, sanitization, encoding, or type casting could
neutralize the propagation. Following this analysis, we generate an
illustrated hop-by-hop breakdown of the data flow. \emph{iii) Document
  Export}, in which all data flow results, verification outcomes, and
a structured vulnerability report are exported for final human expert
review.


\section{Experiments}

\begin{table}[t]
\centering
\resizebox{\linewidth}{!}{%
\begin{tabular}{l|rrrrr}
\Xhline{1.25pt}
\multicolumn{1}{l|}{Repository} & Stars & Commits & Forks & LoC & Issues \\ \hline
PublicCMS & 2.1k  & 3,479   & 828   & 335k & 1  \\
JeecgBoot & \textbf{44.9k} & 2,137   & \textbf{15.7k} & 642k & 19 \\
Ruoyi     & 7.9k  & 1,942   & 2.2k  & 124k & 16 \\
JSPWiki   & 111   & 9,742   & 102   & 161k & 12 \\
DataGear  & 1.6k  & 4,574   & 367   & 190k & 1  \\
Yudao-Cloud & 18.4k & 2,614   & 4.6k  & 230k & 1  \\
KeyCloak    & 32k   & \textbf{29,831}  & 7.9k  & \textbf{841k} & \textbf{2.1k} \\
\Xhline{1.1pt}
\end{tabular}
}
\caption{Details of the evaluation codebases. This table presents the GitHub stars, forks, commits, Lines of Code (LoC), and the number of open issues for each repository.}
\label{tab:repository}
\end{table}

\subsection{Settings}

\noindent
\textbf{Evaluation Environments.}
We take Java codebases vulnerability detection as our evaluation example, and our method can be used in other languages by implementing new dependency analysis agents and new CodeQL query templates.
We evaluate the SAST efficiency and effectiveness of Argus on modern
and industrial-level codebases. Specifically, we select seven popular
open-source codebases, including PublicCMS~\cite{publiccms},
JeecgBoot~\cite{jeecgboot}, Rouyi~\cite{rouyi},
JSPWiki~\cite{jspwiki}, DataGear~\cite{datagear}, Yudao~\cite{yudao},
and KeyCloak~\cite{keycloak}, as detailed in Table \ref{tab:repository}.
All seven codebases contain at least 100,000 lines of code,
with almost 10k GitHub stars and commits, demonstrating that they are
practical and suitable for evaluating realistic vulnerability
detection performance.

\noindent
\textbf{Baselines \& Metrics.}
We employ IRIS~\cite{iris}, the only Java-oriented vulnerability
detection framework, and CodeQL~\cite{Codeql}, the well-known SAST
tool, as our evaluation baselines. Regarding metrics, we utilize the
detected counts of vulnerabilities and sinks to reflect the
effectiveness of methods, and measure token consumption for
computational efficiency.


\subsection{Vulnerability Detection}

We begin our realistic vulnerability detection evaluation with sink
discovery, and then proceed to an end-to-end SAST evaluation.

\begin{table*}[t]
\centering
\resizebox{0.89\linewidth}{!}{%
\begin{tabular}{l|rrrrrrrr}
\Xhline{1.25pt}
\multicolumn{1}{c|}{\multirow{2}{*}{Repository}} & \multicolumn{2}{c}{CodeQL} & \multicolumn{2}{c}{IRIS} & \multicolumn{2}{c}{Argus (Claude-Sonnet-3.5)} & \multicolumn{2}{c}{Argus (Claude-Sonnet-4.5)} \\ \cline{2-9} 
\multicolumn{1}{c|}{} & \multicolumn{1}{r}{\# Sinks} & \multicolumn{1}{r}{\# Vuln.} & \multicolumn{1}{r}{\# Sinks} & \multicolumn{1}{r}{\# Vuln.} & \multicolumn{1}{r}{\# Sinks} & \multicolumn{1}{r}{\# Vuln.} & \multicolumn{1}{r}{\# Sinks} & \multicolumn{1}{r}{\# Vuln.} \\ \hline
PublicCMS & - & 0 & 5840 & 0 & 8 & 6 & 8 & 6 \\
JeecgBoot & - & 0 & 8350 & 0 & 9 & 3 & 9 & 3 \\
Ruoyi & - & 0 & 980& 0 & 3 & 3 & 3 & 3 \\
JSPWiki & - & 0 & 1828 & 1 & 7 & 5 & 7 & 5 \\
DataGear & - & 0 & 237 & 0 & 11 & 3 & 11 & 3 \\
Yudao-Cloud & - & 0 & 2184 & 0 & 9 & 1 & 9 & 1 \\
KeyCloak & - &  0& 13962 & 0 & 19 & 2 & 19 & 2 \\ \Xhline{1.1pt}
\end{tabular}%
}
\caption{End-to-End Vulnerability Detection Evaluation. Argus significantly outperforms CodeQL in traditional vulnerability detection. IRIS fails in realistic SAST scenarios due to the lack of required CWE information.}
\label{tab:e2e}
\end{table*}

\subsubsection{Token Consumption \& Vulnerability Detection Evaluation}
We first compare the LLM API token consumption with the corresponding
number of discovered harmful vulnerabilities. As shown in Table
\ref{tab:consumption}, our token consumption is affordable.
Benefiting from the proposed RAG-enhanced
mechanism, Argus incurs a little bit higher token consumption
compared to the baselines, while identifying much more potential vulnerabilities in
realistic codebases.  As these codebases are actively updated,
in most cases the baseline fail to detect any vulnerabilities, leading to
significant false positives in SAST. 

\begin{table}[t]
\centering
\resizebox{0.86\linewidth}{!}{%
\begin{tabular}{l|rrll}
\Xhline{1.25pt}
\multicolumn{1}{c|}{\multirow{2}{*}{Repository}} & \multicolumn{2}{c}{Argus} & \multicolumn{2}{c}{IRIS} \\ \cline{2-5} 
\multicolumn{1}{l|}{} & \multicolumn{1}{c}{Tokens} & \multicolumn{1}{c}{Vuln.} & \multicolumn{1}{c}{Tokens} & \multicolumn{1}{c}{Vuln.} \\ \hline
PublicCMS & 194,538 & 6 & 35,500 & 0  \\
JeecgBoot & 254,392 & 3 & 42,400 & 0 \\
Ruoyi & 32,836 & 3 & 22,300 & 0 \\
JSPWiki & 67,605 & 5 & 24,638 &1  \\
DataGear & 9,301 & 3 & 20,300 & 0 \\
Yudao-Cloud & 85,661 & 1 & 25,600 & 0 \\
KeyCloak & 442,095 & 2 & 57,641 &0  \\ \Xhline{1.1pt}
\end{tabular}%
}
\caption{LLM API token consumption and vuln. discovery comparison.}
\label{tab:consumption}
\end{table}

\subsubsection{End-To-End Evaluation}
We further compare the automatic vulnerability detection capabilities of
Argus with our baselines in realistic end-to-end scenarios.
Specifically, we count the number of sinks and vulnerabilities for CodeQL,
IRIS, and our Argus. For Argus, we employ two versions of the LLM APIs to
test the sensitivity of the framework to different LLMs. The results are shown
in Table \ref{tab:e2e}.

As shown in Table \ref{tab:e2e}, Argus achieves a consistently higher
number of sinks compared with all baselines, and a significantly higher
number of detected vulnerabilities, indicating substantial progress from
traditional SAST tools and LLM-assisted analysis to LLM-centered
automatic and end-to-end vulnerability detection.

Table \ref{tab:e2e} can also be regarded as a preliminary ablation study.
When comparing the sink numbers from Argus and CodeQL, we can observe
the contribution of extra sinks proposed by our RAG and POC agents. We can infer
that Argus is capable of mining more ``stealthy'' and unknown
sinks than the static CodeQL. Moreover, when we compare the final
reported vulnerabilities, surprisingly, a significant improvement can be
observed, typically increasing from zero to nearly ten vulnerabilities.
This demonstrates that our \emph{Re}$^{3}$ can effectively trace all
potential data flows towards the target sinks.

\begin{table}[]
\centering
\resizebox{0.94\linewidth}{!}{%
\begin{tabular}{l|cc}
\Xhline{1.25pt}
\# Vuln./\# Sinks  &  Bare Argus & Extra with RAG   \\
\hline
PublicCMS   & 5/7                       &  1/1                               \\
JeecgBoot   & 3/9                       &  0/0                               \\
Ruoyi       & 3/3                       &  0/0                               \\
JSPWiki     & 5/7                        & 0/0                                \\
DataGear    & 3/10                          &  3/1                               \\
Yudao-Cloud & 0/0                          &  0/0                               \\
KeyCloak    & 2/17                          &  0/2  \\             
\Xhline{1.1pt}
\end{tabular}
}
\caption{Vulnerabilities resulting from original sink detection and extra sinks detected by RAG agents.}
\label{tab:abl}
\end{table}

If we further track which detected sinks these vulnerabilities originate from,
we can obtain the ablation results shown in Table \ref{tab:abl}.
As indicated in Table \ref{tab:abl}, sinks found by our
RAG agents result in significantly more vulnerabilities compared with the sinks
found by static tools (i.e., bare Argus), indicating that the extra sinks derived from
the community are more valuable for detection.

\section{Case Studies}


\subsection{Case I: How Does Argus ``Recycle'' Previous Outdated Vulnerabilities?}

\begin{figure}
\includegraphics[width=\linewidth]{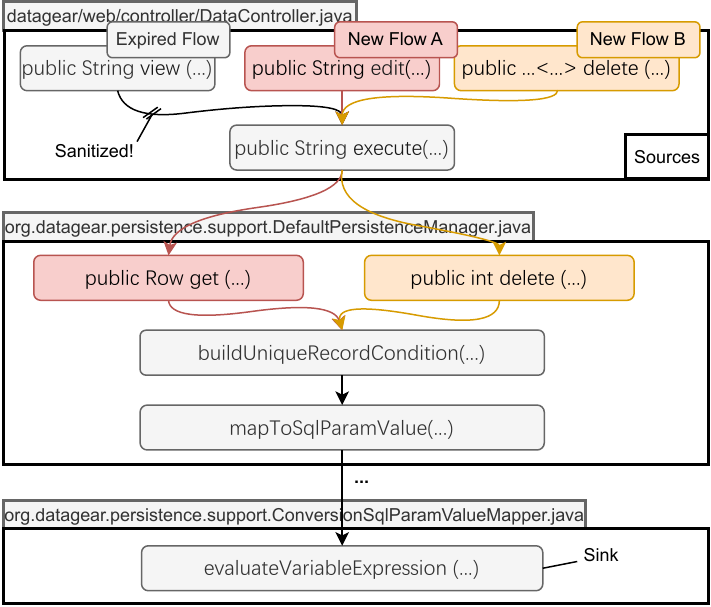}
\caption{Examples of vulnerable data flows in DataGear detected by Argus.}
\label{fig:case1}
\end{figure}

Our first case study is derived from the experiments scanning the
\texttt{DataGear} repository. As shown in Figure \ref{fig:case1},
we observe that the RAG agents in Argus first retrieve a previous
vulnerability named \texttt{CVE-2024-37759}. Through this, it identifies
a sink \texttt{evaluateVariableExpression} at
\texttt{org.datagear...Conversion...ValueMapper}. While
the original vulnerable data flow has been fixed, our
\emph{Re}$^{3}$ identifies two new data flows that
pass our manual review, i.e., they are still exploitable and have not been
discovered before. Moreover, by analyzing the attributes of the new
vulnerabilities, we find several interesting points: \emph{I) The new
  data flows reuse most of the access paths of the previous fixed
  vulnerabilities}, suggesting that reviewing vulnerabilities
heuristically by human experts would lead to high false positives as
many other candidate paths could be neglected. \emph{II) Argus can
  provide more comprehensive vulnerability scanning} via LLM reasoning
  and multi-agent collaboration. \emph{III) Even those vulnerabilities
  which have been fixed remain valuable,} as they highlight the necessity for
more in-depth static analysis regarding the sanitization of injected commands.

\subsection{Case II: Even Incorrect Sinks Can Inspire Argus to Search for
  New Vulnerabilities}

Our second case is derived from experiments on \texttt{PublicCMS},
where our RAG agents discover a CVE (\texttt{CVE-2025-31672}) related to its dependency
\texttt{org.apache.poi:poi-ooxml}. However, the POC agents suspect
that the sink could be \texttt{DocumentBuilderFactory.newInstance} or
\texttt{com...Basic...TypeValidator.builder}, neither of which is
the correct sink capable of triggering XSS or XXE injection attacks.
This discrepancy arises because the two suspected sinks are relevant to the
dangerous behaviors associated with this CVE but are not the exact sinks.

Consequently, when the previous agents provide the sink
\texttt{newInstance}, the forward search of CodeQL cannot identify
any explicit connection from sources to it, leading to a failed
vulnerability detection. However, in the second step of our \emph{Re}$^{3}$,
the true sink \texttt{DocToHtmlUtils.excelToHtml}, an upstream
node of \texttt{newInstance} marked as a new potential sink by our
method, is introduced for the data flow search, thus identifying
the correct vulnerability.
This case demonstrates that the initial incorrect sinks play the role of
effective \emph{seeds}, which can successfully guide the discovery of
the true sinks.


\section{Conclusion}

In this paper, we explore how to construct an LLM-agent-centered
static application security testing method for the vulnerability
detection of modern software. Unlike previous studies, which
merely employ LLMs as a component for specific vulnerability
detection tasks, we reorganize the entire detection procedure with our
new framework, Argus. As a multi-agent system incorporating RAG
techniques, Argus effectively collects existing information about
the repository and dependencies, and employs a new reasoning procedure for
vulnerability verification and data flow mining. Extensive
experiments on realistic open-source codebases demonstrate that Argus
can automatically mine significantly more new zero-day vulnerabilities
compared to existing methods.

\section{Limitations and Future Work}
While our method achieves superior performance in vulnerability detection, several limitations remain in this study. On the one hand, Argus focuses exclusively on static taint analysis and does not incorporate dynamic vulnerability detection techniques, such as fuzzing. Integrating Argus with advanced fuzzing strategies could enable more comprehensive and powerful vulnerability detection in the future. On the other hand, despite our comprehensive agentic system design, numerous LLM opportunities exist to further enhance the effectiveness of the SAST system. For instance, the multi-agent framework could be fine-tuned using reinforcement learning in a vulnerability detection environment, with rewards based on sink discovery accuracy and data flow completeness.

\section*{Ethical Considerations}
Our Argus, leveraging large language models for vulnerability detection, was developed with the primary objective of bolstering cybersecurity defenses. Although there is a slight risk that it could be exploited by malicious actors to uncover vulnerabilities, its predominant application is anticipated in conducting thorough security audits of code repositories, thereby substantially mitigating the likelihood of successful cyber attacks. Moreover, the disclosure of zero-day vulnerabilities underscores the potential for this concise and lightweight approach to be broadly adopted for protective purposes, outweighing any potential misuse.

\bibliography{custom}

\clearpage
\appendix

\begin{figure}[t]
    \centering
    \includegraphics[width=0.98\linewidth]{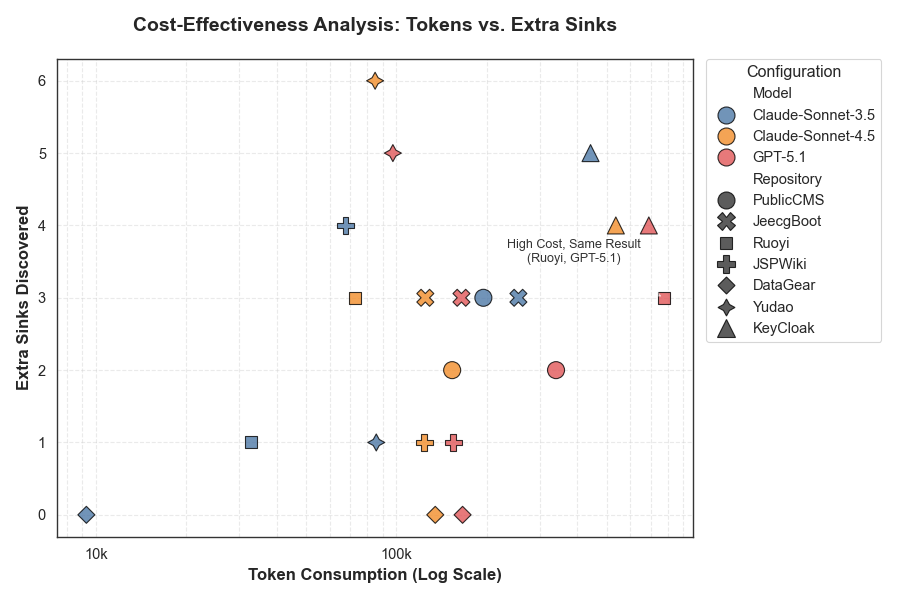}
    \caption{Token consumption and sink discovery performance trade-off among different Argus backbones and codebases.}
    \label{fig:xxx}
\end{figure}

\section{Implementation Details}\label{sec:impl-detail}

We employ Claude-3.5-Sonnet and Claude-4.5-Sonnet as the LLM backbone for our agents and conduct all vulnerability detection experiments on standard CPU servers equipped with 512\,GB of RAM. Argus requires no local GPU resources, as the reasoning workload is entirely handled by remote LLM inference servers. Under this configuration, we measure both token consumption and execution time. For a typical code repository comprising tens of thousands of lines, Argus completes the full detection process in approximately 0.44 hours on average, with an average cost of \$2.54.
To evaluate detected vulnerabilities in our method, we introduce three quantitative criteria:
\emph{i)} Relevance score $\alpha_{r}$, initialized
to 0.5; it increases by 0.4 if the vulnerability description contains
speculative keywords such as potential'' or early'', and gains an
additional 0.1 if it mentions explicit security terms such as
``vulnerability''; however, it decreases by 0.1 if the issue is
already marked with a CVE. \emph{ii)} Credibility score $\alpha_{c}$,
  defined as $\alpha_{c} = 0.3 + \min(N_{c} \times 0.05, 0.3)$, where
  $N_{c}$ is the number of comments on the issue. \emph{iii)} Content
    quality score $\alpha_{q}$, which evaluates the description based
    on length, technical depth, specificity of impact, presence of
    code examples, and proposed solutions.

\begin{figure*}[h]
    \centering
    \includegraphics[width=0.98\linewidth]{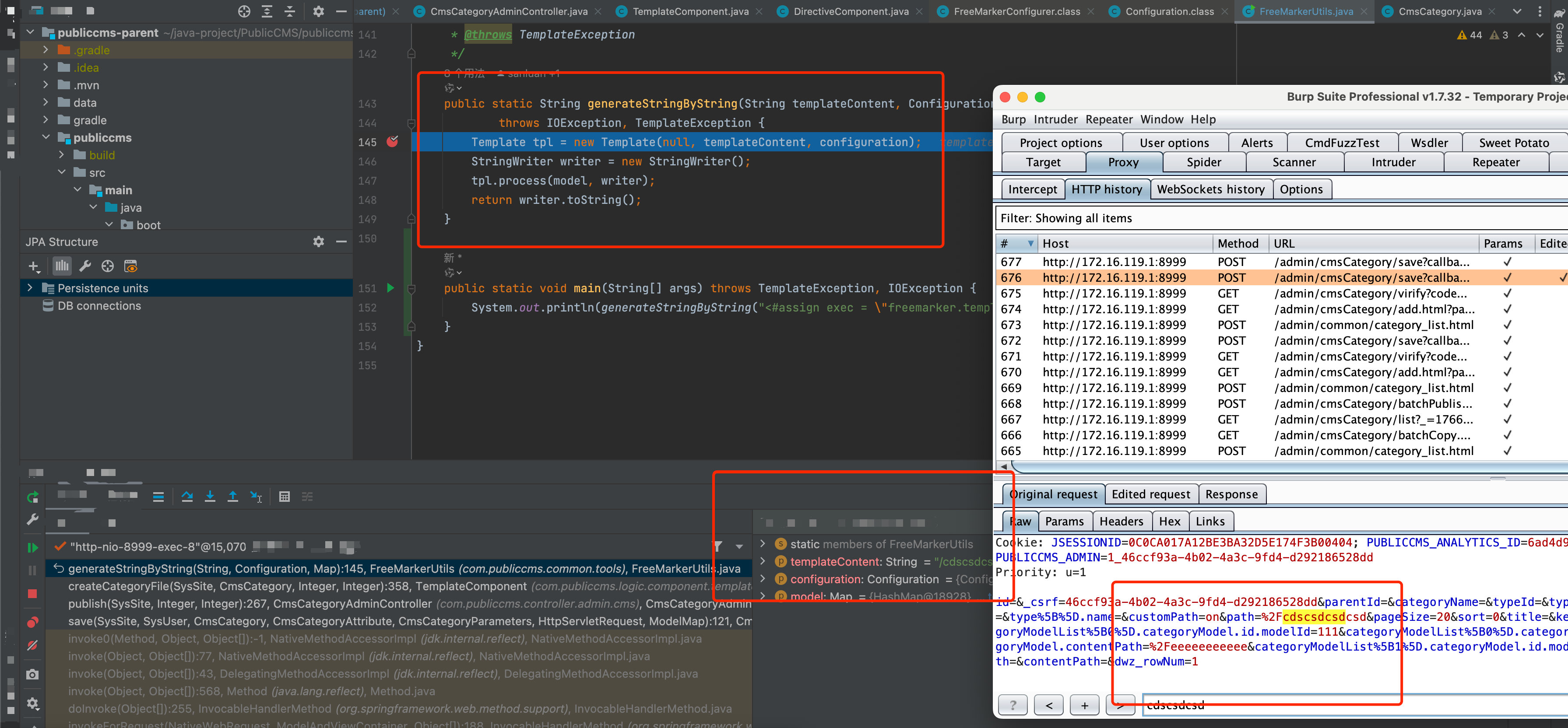}
    \caption{Screenshot of one of our realistic attack examples injected via our zero-day vulnerability discovered with Argus. The repository is PublicCMS.}
    \label{fig:xxx}
\end{figure*}

\section{Supplemental Materials}

\begin{figure*}[htbp]
    \centering
    \includegraphics[width=0.98\textwidth]{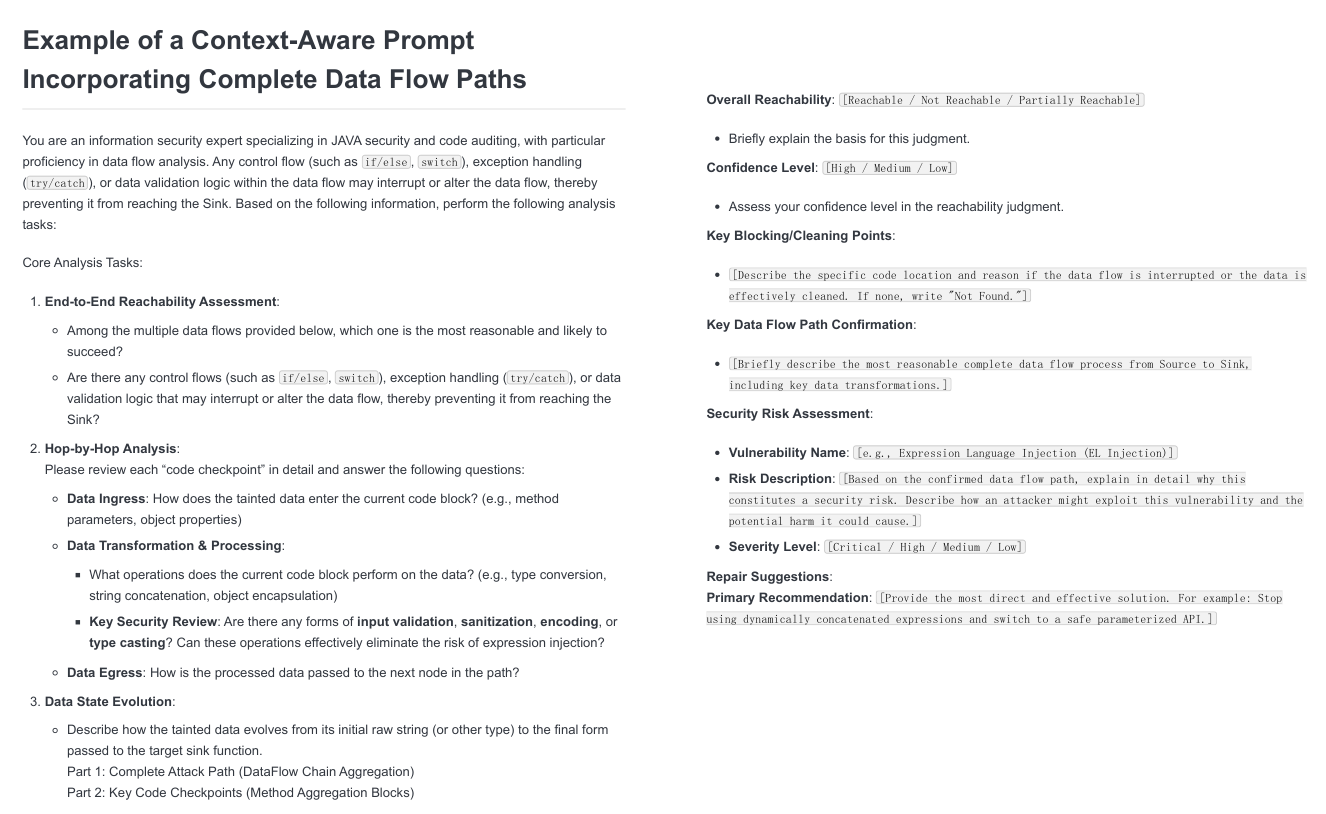}
    \caption{An example prompt of our vulnerability review agent in Argus.}
    \label{fig:xxx}
\end{figure*}

\begin{figure*}[htbp]
    \centering
    \includegraphics[width=0.98\textwidth]{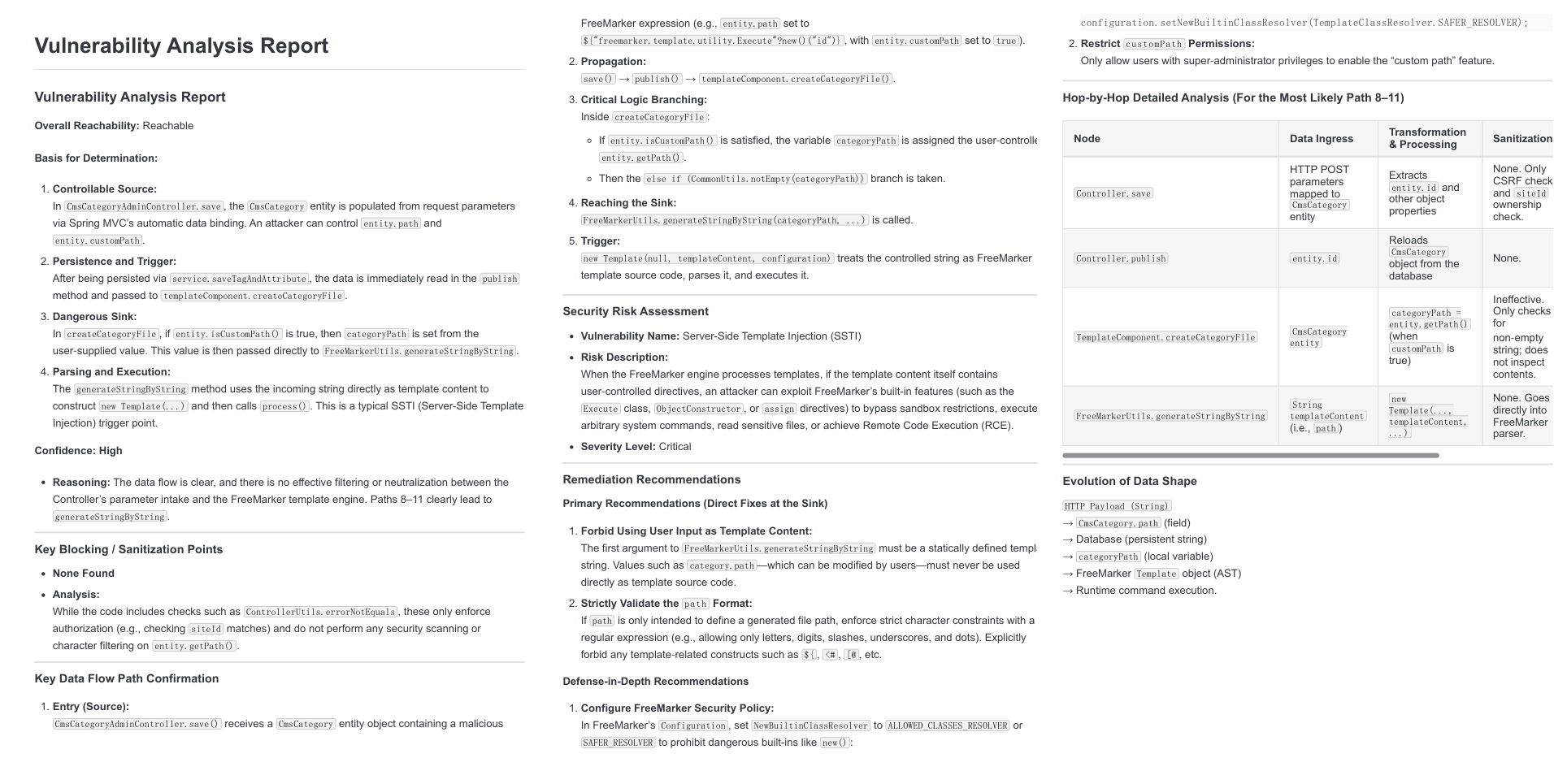}
    \caption{The final vulnerability analysis report exported by Argus.}
    \label{fig:xxx}
\end{figure*}


\end{document}